\documentclass[preprint,showpacs]{revtex4}

\usepackage{graphicx}
\usepackage{dcolumn}
\usepackage{bm}
\usepackage{amsmath}
\usepackage[dvips]{color}

\definecolor{gris25}{gray}{0.5}
\begin{document}

\preprint{Aubry et al.}

\title[Multiple scattering in weakly inhomogeneous media]{Multiple scattering of ultrasound in weakly inhomogeneous media: Application to human soft tissues}

\author{Alexandre Aubry, Arnaud Derode}

\affiliation{Institut Langevin, ESPCI ParisTech \\
CNRS UMR 7587, Universit\'e Denis Diderot (Paris VII)\\
10 rue Vauquelin, 75005 Paris, France}

%

\date{\today}

\begin{abstract}
Waves scattered by a weakly inhomogeneous random medium contain a predominant single scattering contribution as well as a multiple scattering contribution which is usually neglected, especially for imaging purposes. A method based on random matrix theory is proposed to separate the single and multiple scattering contributions. The experimental set up uses an array of sources/receivers placed in front of the medium. The impulse responses between every couple of transducers are measured and form a matrix. Single-scattering contributions are shown to exhibit a deterministic coherence along the antidiagonals of the array response matrix, whatever the distribution of inhomogeneities. This property is taken advantage of to discriminate single from multiple-scattered waves. This allows one to evaluate the absorption losses and the scattering losses separately, by comparing the multiple scattering intensity with a radiative transfer model. Moreover, the relative contribution of multiple scattering in the backscattered wave can be estimated, which serves as a validity test for the Born approximation. Experimental results are presented with ultrasonic waves in the MHz range, on a synthetic sample (agar-gelatine gel) as well as on breast tissues. Interestingly, the multiple scattering contribution is found to be far from negligible in the breast around 4.3 MHz.

\copyright{Copyright 2011 Acoustical Society of America. This article may be downloaded for personal use only. Any other use requires prior permission of the author and the Acoustical Society of America. The following article appeared in J. Acoust. Soc. Am. \textbf{129}, 255-233 (2011), and may be found at \href{http://link.aip.org/link/?JAS/129/225}{http://link.aip.org/link/?JAS/129/225}}
\end{abstract}

%

\maketitle

\section{\label{sec:intro}Introduction}

\begin{figure}[htbp] 
\includegraphics{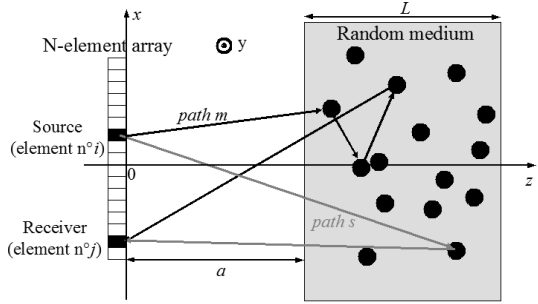}
\caption{\label{fig:setup}Experimental setup: a 125-element linear array is placed in front of a random medium at a distance $a$. The whole setup is immersed in a water tank. 
}
\end{figure}

Standard imaging techniques, such as ultrasonic echography \cite{angelsen}, radar \cite{stergiopoulos} or optical coherence tomography \cite{Schmitt}, are based on the same principle. One or several source(s) emit(s) a wave into the medium to be imaged. It is reflected by the inhomogeneities of the medium and the backscattered wave is measured by the same or other sensor(s). It contains two contributions:

\noindent $\bullet$ A single scattering contribution (path $s$ in Fig.\ref{fig:setup}): the incident wave undergoes only one scattering event before coming back to the sensor(s). This is the contribution which is used in imaging because there is a direct relation between the arrival time $t$ of the echo and the distance $d$ between the sensors and the scatterer, $t=2d/c$ ($c$ is the sound velocity). An image of the medium reflectivity can be built from measured signals.

\noindent $\bullet$ A multiple scattering contribution (path $m$ in Fig.\ref{fig:setup}): the wave undergoes several scattering events before reaching the sensor. Multiple scattering occurs when scatterers are strongly diffusive and/or highly concentrated. There is no correspondence between the arrival time $t$ and the position of a scatterer. Thus, classical imaging fails in multiple scattering media \cite{bly,bordier,karamata,karamata2}.

{Standard imaging techniques rely on the single-scattering assumption (first Born approximation). However, there is no such thing as a purely single scattering medium. A multiple scattering contribution always exists, albeit negligible compared to single scattering. Naturally for imaging purposes, one tries to reduce the influence of multiple scattering, for instance by choosing an appropriate frequency domain where multiple scattering is not too strong \cite{aubry4}. Focused beamforming with an array of transducers, or more generally synthetic aperture techniques  \cite{stergiopoulos}, are also a way to enhance the single scattering contribution. It should be noted that even though multiple scattering is considered as the enemy of classical imaging techniques, studying it may bring additional information about the scattering structure. Indeed, a wave undergoing multiple scattering can be thought of as a random walker \cite{rossum}, with two essential parameters: the elastic mean-free path $l_e$ and the diffusion constant $D$. Measuring these parameters is a way to characterize the microarchitecture of the scattering medium \cite{rytov,rytov2}. Yet in weakly inhomogeneous media where the first Born approximation is reasonably valid (especially human soft tissues probed by ultrasound in the MHz range), it is a challenge to study multiple scattering parameters because of the predominance of single scattering.}

{Recently, an original technique has been proposed to separate the single-scattered echo of a target drowned in a predominant multiple scattering background \cite{aubry5,aubry4}. The method was based on a matrix approach. It has been successfully applied to target detection and imaging in highly scattering media. In this paper, we are also interested in discriminating single-scattering and multiple-scattering contributions from the total response of an unknown medium, based on matrix properties. However, the present approach is different from earlier works \cite{aubry5,aubry4}, both in terms of method and applications. The situation we consider here is exactly the opposite: we want to extract the multiple scattering contribution from predominantly single-scattered waves in a weakly scattering media. Moreover, the method is based on a singular value decomposition applied to the antidiagonals of the array response matrix, and not to the array response matrix itself. The distinction between single and multiple scattering subspaces is then performed using random matrix theory \cite{tulino,sengupta}, as it will be detailed in the next sections.}

The interest of this work is twofold. First, once single and multiple scattering contributions are isolated, the proportion of multiple scattering within the wave response of the medium can be evaluated. This figure can be used as an indicator for the validity of the single-scattering (Born) approximation, which is the basis of classical imaging techniques. Note however that the purpose of this study is not to improve imaging of weakly scattering media. The second interest of this work is to provide a new tool for the characterization of weakly scattering media. More precisely, we will show how the multiple scattering contribution, once it is isolated, can be taken advantage of in order to estimate the scattering mean-free path $l_e$ independently from the absorption mean-free path $l_a$, thus discriminating absorption and scattering losses. 

The experimental results we present were obtained with pulsed ultrasonic waves firstly in a synthetic medium (agar-gelatine gel) around 3 MHz then in breast tissues around 4.3 MHz, but the principle of the technique can be applied to all fields of wave physics (\textit{e.g} seismology, electromagnetism, acoustics \textit{etc.}) for which the multi-element array technology is available and provides time-resolved measurements of the amplitude and the phase of the wave field.

\section{\label{sec:exp_num}Transducers' array configuration}
We use a N-element ultrasonic array (here, $N = 125$). The array is placed at a distance $a$ from the random scattering sample under investigation (see Fig.~\ref{fig:setup}). The first step of the experiment consists in measuring the inter-element matrix. A sinusoidal burst of length $\delta t$, at the central frequency $f_c$, is emitted from transducer $i$ into the scattering medium. Typical values here are $\delta t \sim 1$ $\mu$s and $f_c \sim 3$ MHz. The backscattered wave is recorded with the $N$ transducers of the same array, which yields a set of impulse responses $h_{ij}(t)$ ($j=1,...,N$ denotes the receiver index). The operation is repeated for the $N$ emitting transducers. The responses $h_{ij}(t)$ form the $N \times N$ impulse response matrix $\mathbf{H}(t)$. Because of reciprocity, $h_{ij}(t)=h_{ji}(t)$ and $\mathbf{H}(t)$ is symmetric. A short-time Fourier analysis of the impulse response matrix $\mathbf{H}$ is performed. The time signals $h_{ij}(t)$ are truncated into $\Delta t$-long overlapping windows : $k_{ij}(T,t)=h_{ij}(T-t)W_R(t)$ with $W_R(t)=1 \; \text{for} \; t\in[-\Delta t / 2 \; , \; \Delta t / 2]$, $W_R(t)=0$ anywhere else. The value of $\Delta t$ is chosen so that signals associated with the same scattering event(s) within the medium arrive in the same time window \cite{aubry3}. {Typical values here are $\Delta t \sim 10$ $\mu$s.} A Fourier analysis of $\mathbf{K}(T,t)$ is achieved by means of a discrete Fourier transform. A response matrix $\mathbf{K}(T,f)$ is finally obtained at each time $T$ and frequency $f$.
The single and multiple scattering contributions can now be discriminated with the help of a matrix manipulation.

\section{\label{sec:coherence_ss}The signature of single scattering}

\begin{figure}[htbp] 
\includegraphics{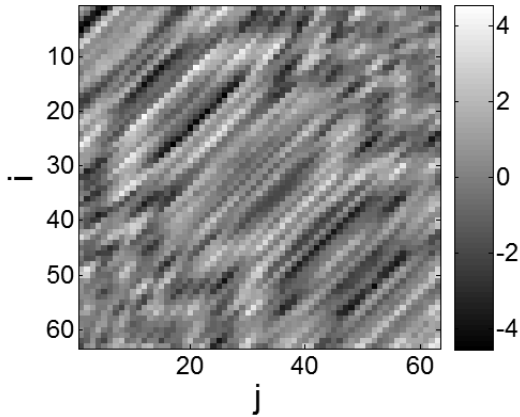}
\caption{\label{fig:fig2}Real part of matrix $\mathbf{K}$ obtained in a gel (5\% gelatine, 3\% agar-agar) at time $T=114$ $\mu$s and frequency $f=$3.05 MHz. The source-sample distance was $a=50$ mm. }
\end{figure}

When studying the array response matrices $\mathbf{K}(T,f)$, the predominance of single scattering manifests itself by the presence of a long-range deterministic coherence along the antidiagonals of the matrix, whatever the distribution of scatterers \cite{aubry4,aubry5,aubry3}. As an example, Fig.\ref{fig:fig2} shows the real part of one of the matrices $\mathbf{K}$, in the case of a synthetic medium (Agar-gelatine gel) which is enough weakly scattering for the Born approximation to be valid. Even if the inhomogeneities are randomly distributed, $\mathbf{K}$ obviously exhibits some kind of coherence along its antidiagonals (i.e., for matrix elements $k_{ij}$ such that $i+j =constant$). This coherence is a typical signature of single scattering, and it vanishes when multiple scattering dominates. This has been thoroughly explained in \cite{aubry4,aubry5,aubry3}, we briefly recall the main argument here.

Generally, the $k_{ij}(T,f)$ can be written as the sum of single and multiple scattering contributions:
\begin{equation}
\label{eqn:ss_ms}
k_{ij}(T,f)=k_{ij}^S(T,f)+k_{ij}^M(T,f)
\end{equation}
Under the paraxial approximation, {the distance between the origin $(0,0)$ and an observer $(x,R)$ located slightly off-axis ($x<<R$) is $\sqrt{R^2+x^2}=R+{x^2}/({2R})$.} As a result, the phase shift undergone by a wave travelling from a source with coordinates $(0,x_i)$, scattered by a point $(X_d,R)$ and received in the plane of the source at $(0,x_j)$ reads :
\begin{equation*}
\exp \left [2jkR \right ]\exp \left [ j k \frac{ \left (x_i -X_d \right )^2}{2R} \right ] \exp \left [ j k \frac{ \left (x_j -X_d \right )^2}{2R} \right ]
\end{equation*}
with $k$ the wavenumber. The quadratic phase terms can be factorized since
\begin{equation}
\label{factor}
\left (x_i -X_d \right )^2 + \left (x_j -X_d \right )^2= \frac{\left (x_i - x_j \right )^2}{2} + \frac{\left (x_i + x_j - 2 X_d \right )^2}{2}
\end{equation}

Consider an ensemble of scatterers randomly distributed. As long as only the first and the last scattering of every scattering path are identical (which is naturally the case, if only single scattering takes place) the coefficients of the array response matrix {at time $T$ and frequency $f$} will be proportional to:
\begin{equation}
\label{eqn:eq_ss_signal}
k_{ij}^S(T,f) \propto  {\exp \left ( j 2kR \right )}\sum_{d=1}^{N_d} A_d \exp \left [ j k \frac{ \left (x_i -X_d \right )^2}{2R} \right ] \exp \left [ j k \frac{ \left (x_j -X_d \right )^2}{2R} \right ]
\end{equation}
{with $R=cT/2$ and} $N_d$ the number of scatterers {contained in the isochronous volume}. $A_d$ depends on the reflectivity of the scatterer. {$A_d$ and $X_d$ are random variables} so $k_{ij}^S$ is itself random. {Interestingly, applying the factorization of Eq.\ref{factor} to Eq.\ref{eqn:eq_ss_signal}, a deterministic relation arises along the antidiagonals of $\mathbf{K^S}$:}
\begin{equation}
\label{eqn:eq_ss_signal_3}
\beta_m = \frac{k_{i-m,i+m}(T,f)}{k_{ii}(T,f)} =  \exp \left [ j k \frac{ \left (m \delta x \right )^2}{R} \right ]
\end{equation}
with $\delta x$ the array pitch and $2m \delta x$ denotes the distance between two array elements ($i-m$ and $i+m$) on the same antidiagonal. Eq.\ref{eqn:eq_ss_signal_3} implies that as long as there is only single scattering, there must be a form of coherence, a long-range deterministic relation, between the elements of the array response matrix, whatever the realisation of disorder.
On the contrary, when multiple scattering occurs (except for recurrent scattering paths \cite{wiersma}, but this contribution is negligible in weakly scattering media), the elements $k_{ij}^M$ cannot be factorized, and there is no such long-range deterministic coherence \cite{aubry4,aubry5,aubry3}.

\section{\label{sec:separation_ss_ms}Separation of single and multiple scattering}
The key to separate single ($\mathbf{K^S}$) and multiple ($\mathbf{K^M}$) scattering contributions is the particular coherence of $\mathbf{K^S}$ along its antidiagonals. In previous works \cite{aubry4,aubry5}, $\mathbf{K^S}$ was extracted from $\mathbf{K}$ by projecting the antidiagonals of $\mathbf{K}$ along the vector $[ \beta_m ]$ of Eq.\ref{eqn:eq_ss_signal_3}. But the simple form taken by Eq.\ref{eqn:eq_ss_signal_3} results from  a series of assumptions (paraxial approximation, pointlike array elements and scatterers) which do not all apply to our experimental configuration. In order to separate $\mathbf{K^S}$ and $\mathbf{K^M}$, the method proposed in this paper is much less restrictive. We do not assume that Eq.\ref{eqn:eq_ss_signal_3} exactly applies; we only assume that because of single scattering there must be a deterministic coherence between the antidiagonal elements of $\mathbf{K^S}$, but we do not suppose we know its exact form. 

Under these conditions, the separation between $\mathbf{K^S}$ and $\mathbf{K^M}$ will essentially rely on a singular value decomposition (SVD) {of the antidiagonals of $\mathbf{K}$}. This separation is a three-step process:

\noindent $\bullet$ Rotation of each matrix $\mathbf{K}$ and construction of two sub-matrices $\mathbf{A_1}$ and $\mathbf{A_2}$.

\noindent $\bullet$ Filtering of matrices $\mathbf{A_r}$ $(r=1,2)$. $\mathbf{A_r}$  is decomposed as the sum of two matrices: $\mathbf{A_r}=\mathbf{A_{r}^S}+\mathbf{A_{r}^M}$, where $\mathbf{A_{r}^S}$ and $\mathbf{A_r^M}$ contain respectively the single and multiple scattering signals. 

\noindent $\bullet$ Construction from $\mathbf{A_r^{S}}$ and $\mathbf{A_r^{M}}$ of the single and multiple scattering matrices $\mathbf{K^{S}}$ and $\mathbf{K^{M}}$.

{The first and third steps (rotation of data) have already been presented in previous works \cite{aubry4,aubry5} and will be briefly recalled in Sec.\ref{subsec:rot_anti} and Sec.\ref{subsec:buiding_Kf}. On the contrary, the second step (SVD of antidiagonals) constitutes the core of the method and differs completely from the previous approach  \cite{aubry4,aubry5}. The corresponding matrix operations are explained in details in Sec.\ref{subsec:filter_anti}.}

\subsection{\label{subsec:rot_anti}First step}
\begin{figure}[htbp] 
\includegraphics{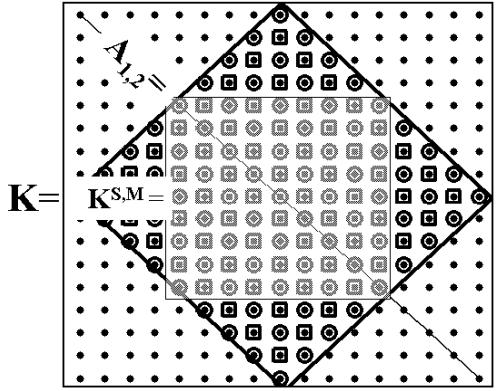}
\caption{\label{fig:rotation} Example of a matrix $\mathbf{K}$ of dimension $N =$ 17. The black points represent the elements $k_{ij}$ of $\mathbf{K}$. The antidiagonals of $\mathbf{K}$ are the columns of matrices $\mathbf{A_1}$ and $\mathbf{A_2}$. Circles and squares represent the elements of $\mathbf{A_1}$ and $\mathbf{A_2}$, respectively. Once single and multiple scattering contributions are separated, the final matrices $\mathbf{K^S}$ and $\mathbf{K^M}$ have $(2M - 1) \times (2M - 1)$ elements (central square).}
\end{figure}

A rotation of matrix data is achieved as depicted in Fig.\ref{fig:rotation}. It consists in building two matrices $\mathbf{A_1}$ and $\mathbf{A_2}$ from matrix $\mathbf{K}=\left [k_{ij} \right ]$:
\begin{eqnarray}
\label{eqn:construction_A1}
\mathbf{A_1}   =  \left [ a_{1uv} \right ]  \,  \mbox{of dimension }(2M-1) \times (2M-1) \mbox{,}\nonumber \\
 \mbox{such that }  a_1[u, v]  = k[u + v -1, v - u + 2M -1] \\
\mathbf{A_2} =  \left [ a_{2uv} \right ]  \,  \mbox{of dimension }(2M-2) \times (2M-2)\mbox{,}\nonumber \\
 \mbox{such that }  a_2[u, v]  = k[u + v , v - u + 2M -1]
\label{eqn:construction_A2}
\end{eqnarray}
where $M=(N+3)/4$. Here $N=125$ and so $M=32$ is an even number.
The columns of matrices $\mathbf{A_1}$ and $\mathbf{A_2}$ correspond to the antidiagonals of $\mathbf{K}$ (see Fig.\ref{fig:rotation}). In the next subsection, we will no longer make the difference between matrices $\mathbf{A_1}$ and $\mathbf{A_2}$ because they are filtered in the same way. They will be called indifferently $\mathbf{A}$. $L$ is the dimension of $\mathbf{A}$. For matrix $\mathbf{A_1}$, $L=2M -1$; for matrix $\mathbf{A_2}$, $L=2M -2$. Because of spatial reciprocity, $\mathbf{K}$ is symmetric ($k_{ij}=k_{ji}$). Thus, $\mathbf{A}$ exhibits also a symmetry: each line of its upper part is identical to a line of its lower part. The symmetry axis is shown as a black line in Fig.\ref{fig:rotation} and corresponds to the diagonal of the matrix $\mathbf{K}$. So, each column of the matrix $\mathbf{A}$ contains only $M$ independent coefficients, even if its dimension $L$ is larger than $M$.

\subsection{\label{subsec:filter_anti}Second step}

$\mathbf{A}$ can be written as a sum of two matrices $\mathbf{A^S}$ and $\mathbf{A^M}$, which correspond respectively to the single and multiple scattering contributions
\begin{equation}
\label{eqn:ss_ms_anti}
\mathbf{A}=\mathbf{A^S}+\mathbf{A^M}
\end{equation}
Contrary to previous works \cite{aubry5,aubry4}, the technique we propose in this paper consists in separating single and multiple scattering by achieving the singular value decomposition (SVD) of the matrix $\mathbf{A}$. The SVD decomposes a matrix into two subspaces: a \textit{signal} subspace (a matrix characterized by an important correlation between its lines and/or columns) and  a \textit{noise} subspace (a random matrix without any correlations between its entries). When the SVD is applied to the matrix $\mathbf{A}$, the \textit{signal} subspace (\textit{i.e.}, the largest singular values) corresponds to $\mathbf{A^S}$ (the single scattering contribution characterized by a long-range correlation along its columns) and the \textit{noise} subspace (\textit{i.e.}, the smallest singular values) corresponds to $\mathbf{A^M}$ (the multiple scattering contribution). 

The SVD of matrix $\mathbf{A}$ is given by
\begin{equation}
\label{eqn:svd_A}
\mathbf{A}=\mathbf{U}\mathbf{\Lambda}\mathbf{V}^{\dag}=\sum_{k=1}^L \lambda_k \mathbf{U}_k  \mathbf{V}_k^{\dag}
\end{equation}
$\mathbf{U}$ and  $\mathbf{V}$ are square unitary matrices of dimension $L$. Their respective columns $\mathbf{U}_k$ and $\mathbf{V}_k$ correspond to the singular vectors associated to the singular value $\lambda_k$. $\mathbf{\Lambda}$ is a square diagonal matrix of dimension $L$, containing the real positive singular values $\lambda_k$ in a decreasing order ($\lambda_1>\lambda_2>...> \lambda_L$). Actually, $\mathbf{A}$ has only $M$ non-zero singular values since it contains only $M$ independent lines, hence Eq.\ref{eqn:svd_A} becomes:
\begin{equation}
\label{eqn:svd_A_2}
\mathbf{A}=\mathbf{U}\mathbf{\Lambda}\mathbf{V}^{\dag}=\sum_{k=1}^M \lambda_k \mathbf{U}_k  \mathbf{V}_k^{\dag}
\end{equation}

The issue is to determine which rank of singular value separates the \textit{signal} subspace (single scattering) from the \textit {noise} subspace (multiple scattering). If Eq.\ref{eqn:eq_ss_signal_3} were strictly true, the single scattering contribution $\mathbf{A^S}$ would be of rank 1 and only the first singular space associated to the first singular value $\lambda_1$ would correspond to the signal subspace. But when assumptions leading to Eq.\ref{eqn:eq_ss_signal_3} do not strictly hold, $\mathbf{A^S}$ is no longer of rank 1 and several singular spaces (associated to the largest singular values) are needed to fully describe the signal subspace. This happens for instance when scatterers are not pointlike or when the paraxial approximation does not hold. We have to define a threshold to discriminate the signal and noise subspaces, with the help of random matrix theory (RMT) \cite{tulino,sengupta}. 

By convention and for the sake of simplicity, the singular values $\lambda_k$ are first normalized by their quadratic mean
\begin{equation}
\label{eqn:sing_val_renorm}
\tilde{\lambda}_k=\frac{\lambda_k}{\sqrt{M^{-1}\sum_{q=1}^M\lambda^{2}_q}}
\end{equation}
For a random matrix of dimension $P \times Q$ (with $1<<P<Q$), whose entries are complex random variables, independently and identically distributed, the probability density function $\rho(\lambda)$ of the normalized singular values $\tilde{\lambda}_k$ is given by \cite{sengupta}
\begin{equation}
\label{eqn:dens_sing_val}
\rho(\lambda)=\frac{1}{\pi\lambda}\sqrt{\left (\lambda_{\mbox{\small max}}^2 -\lambda^2 \right)\left (\lambda^2 -\lambda^2_{\mbox{\small min}} \right)} 
\end{equation}
for $\lambda_{\mbox{\small min}}<\lambda<\lambda_{\mbox{\small max}}$ and 0 otherwise, with
\begin{equation}
\label{eqn:dens_sing_val_2}
\lambda_{\mbox{\small max,min}}=1\pm \sqrt{{P}/{Q}}.
\end{equation}
For random matrices of large dimensions, the singular value spectrum has a bounded support. In our case, $\mathbf{A}$ is a square matrix of dimension $L \times L$. Yet, as it contains only $M$ independent lines, it is equivalent to a rectangular $M \times L$ matrix. If it were truly random (which is expected to be the case of the multiple scattering contribution) its largest singular value should not exceed $\lambda_{\mbox{\small max}}=1.71$. This value is obtained from Eq.\ref{eqn:dens_sing_val_2}, with $P=M=32$ and $Q=L=63$.

It should be noted that rigorously $L$ and $M$ are not large enough for the asymptotic law (Eq.\ref{eqn:dens_sing_val}) to apply. Actually the first singular value $\tilde{\lambda}_1$ obeys a complicated law, known as the Tracy-Widom distribution \cite{johnstone}, which is of unbounded support. The probability for $\tilde{\lambda}_1$ to be larger than $\lambda_{\mbox{\small max}}$ can be computed: it is found to be $\sim $0.08 for $P=M=32$ and $Q=L=63$. The presence of correlations between matrix entries also induce a deviation from Eq.\ref{eqn:dens_sing_val}, as we will see later.  For the sake of simplicity, we admit for now that within an acceptable probability of error, the singular values are upper bounded by $\lambda_{\mbox{\small max}}$.

According to Eq.\ref{eqn:ss_ms_anti}, $\mathbf{A}$ is the sum of a matrix $\mathbf{A^S}$ of rank $p<M$ (associated to single scattering) and a matrix $\mathbf{A^M}$ of rank M (associated to multiple scattering). Sengupta and Mitra \cite{sengupta} have shown that the $(M-p)$ smallest singular values (linked to the noise subspace) exhibit the same distribution as singular values of a random matrix whose size is $(M-p)\times L$. Let $\lambda_{ \mbox{\small max}}^{(q)}$ denote the upper bound of the singular values distribution in the case of a random matrix of dimension $(M-q)\times L$; we have:
\begin{equation}
\label{eqn:dens_sing_val_3_chap3}
\lambda_{\mbox{\small max}}^{(q)}=1+ \sqrt{({M-q})/{L}}
\end{equation}

\begin{figure}[htbp] 
\includegraphics{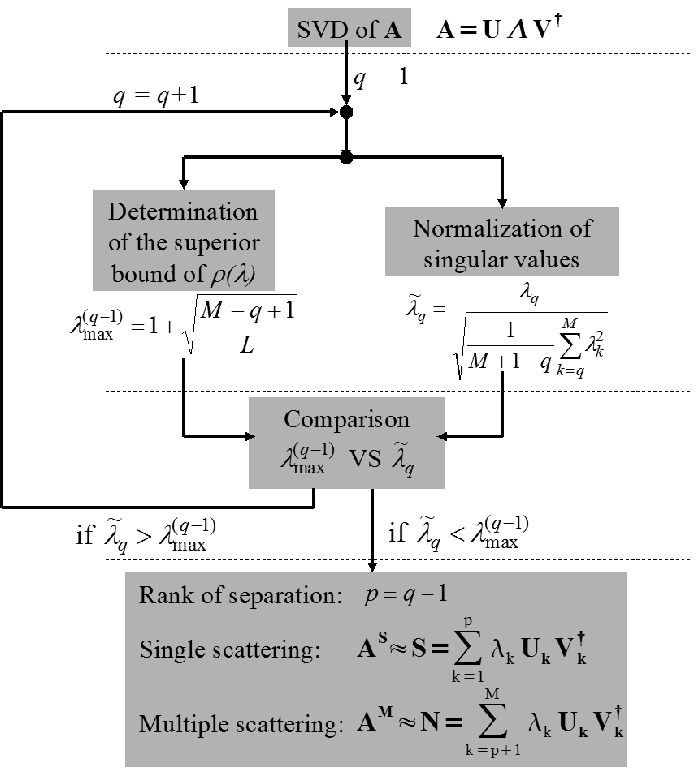}
\caption{\label{fig:fig4}Principle of the separation between the single and multiple scattering contributions.}
\end{figure}

From this property, one can propose a way to separate the signal and noise subspaces of $\mathbf{A}$. We first consider the first singular value $\tilde{\lambda}_1$ upon normalization (Eq.\ref{eqn:sing_val_renorm}). If $\tilde{\lambda}_1$ is larger than $\lambda_{\mbox{\small max}} ^{(0)}$ (Eq.\ref{eqn:dens_sing_val_3_chap3}), it means that the first singular space $\lambda_1 \mathbf{U_1} \mathbf{V^{\dag}_1}$ is associated with the signal subspace. Then, we iterate the process and consider the second singular value. The $\lambda_k$ are once again renormalized, considering only the singular values from $k = 2$:
\begin{equation}
\label{eqn:sing_val_renorm_2}
\tilde{\lambda}_2=\frac{\lambda_2}{\sqrt{(M-1)^{-1}\sum_{k=2}^M\lambda^{2}_k}}
\end{equation}
The threshold value $\lambda_{\mbox{\small max}}$ to consider this time  is the one obtained for a random matrix of size $(M-1)\times L$, \textit{i.e} $\lambda_{\mbox{\small max}} ^{(1)}$ (Eq.\ref{eqn:dens_sing_val_3_chap3}). If $\tilde{\lambda}_2>\lambda_{\mbox{\small max}} ^{(1)}$, the second singular space $\lambda_2 \mathbf{U_2} \mathbf{V_2^{\dag}}$ is also linked to the single scattering contribution, and we iterate once more the process until rank $p+1$ for which $\tilde{\lambda}_{p+1} < \lambda_{\mbox{\small max}} ^{(p)}$. Finally, we obtain a threshold rank $p$ which allows to separate the signal ($\mathbf{S}$) and noise ($\mathbf{N}$) subspaces,
\begin{equation}
\label{eqn:svd_A_3}
\mathbf{S} = \sum_{k=1}^{p} \lambda_k \mathbf{U_k}  \mathbf{V_k^{\dag}} \mbox{ , }\mathbf{N} = \sum_{k=p+1}^M \lambda_k \mathbf{U_k}  \mathbf{V_k^{\dag}}
\end{equation}
Ideally, $\mathbf{S}$ should be devoid of multiple scattering. This is not strictly true, because multiple scattering signals are not strictly orthogonal to the single scattering subspace. Let $\sigma_S^2$ and $\sigma_M^2$ be the power of single and multiple scattering signals. In Appendix A, we show that the typical amplitude of the remaining multiple scattering contribution in $\mathbf{S}$ is $\sigma_M\sqrt{p/2M}$ ($<<\sigma_S$). If we neglect this residual term, we have separated single and multiple scattering contributions: $\mathbf{A^S}\simeq \mathbf{S}$ and $\mathbf{A^M} \simeq \mathbf{N}$. The whole separation process is summarized in Fig.\ref{fig:fig4}. 

{Note that a multiple scattering rate $\gamma$ can be directly measured from the singular values $\lambda_k$ of $\mathbf{A}$. The sum of the square of all the singular values corresponds to the total intensity backscattered by the medium towards the transducers' array . Hence, a multiple scattering rate $\gamma$ can be estimated from the singular values $\lambda_k$ of A:
\begin{equation}
\label{G}
\gamma=\frac{\sum_{k=p+1}^M \lambda_k^2}{\sum_{k=1}^M \lambda_k^2}
\end{equation}
}

Until now, for simplicity we have implicitly assumed that along the antidiagonals of $\mathbf{K^M}$ (or the columns of $\mathbf{A^M}$) the matrix elements are completely decorrelated. However, experimentally short-range correlations may exist between elements, mostly because of mechanical coupling between neighboring transducers and of the coherence length of the diffuse wave-field \cite{aubry5,aubry3}. Correlations between matrix elements can be taken into account theoretically \cite{sengupta}. Consequently, the actual probability density function $\rho(\lambda)$ is more complicated than the simple result of Eq.\ref{eqn:dens_sing_val}, which modifies the upper bound $\lambda_{\mbox{\small max}}^{(q)}$ \cite{aubry5,aubry3}. {In practice, $\lambda_{\mbox{\small max}}^{(q)}$ has to be computed numerically, based on a acceptable probability of error (typically 1\%). The details are given in Appendix B.

This technique of separation is based on the fact that the first singular value exceeds the value $\lambda_{\mbox{\small max}}$, otherwise there is no separation between single and multiple scattering and the whole signal is considered to be associated with multiple scattering. So, this approach is not well suited for strongly diffusive media, $\textit{i.e}$ random media for which the multiple scattering contribution is predominant \cite{aubry5,aubry4}.

\subsection{\label{subsec:buiding_Kf}Third step}
The third step is the reverse of the first one. From $\mathbf{A^S}$ and $\mathbf{A^M}$, two matrices $\mathbf{K^S}$ and $\mathbf{K^M}$, of dimension $(2M-1)\times(2M-1)$, are built(see Fig.\ref{fig:rotation}) with a change of coordinates, back to the original system:

\noindent $\bullet$ if $(i-j)/2$ is an integer,\\ then, $k^{S,M}[i, j]= a_1^{S,M} [(i - j)/ 2 + M,(i + j)/ 2]$

\noindent $\bullet$ if $(i-j)/2$ is not an integer,\\ then, $ k^{S,M}[i, j]= a_2^{S,M} [(i - j -1)/ 2 + M, (i + j -1)/ 2]$

\noindent $\mathbf{K^S}$ contains the single scattering contribution (plus a residual multiple scattering contribution) and $\mathbf{K^M}$ contains the multiple scattering contribution. 

\section{\label{sec:agar_gel}Characterization of a weakly scattering medium}

The experimental set up has already been described in Sec.\ref{sec:exp_num} and is shown in Fig.\ref{fig:setup}. The experiment takes place in a water tank. The ultrasonic array has $N=125$ elements. The emitted signal is a sinusoidal burst of length $\delta t = 2.5$ $\mu$s at the central frequency (3 MHz). The sampling frequency is 20 MHz. Each array element is 0.39 mm in size and the array pitch $\delta x$ is 0.417mm. The source-sample distance is $a=50$ mm. The first random medium of interest is a gel composed of 5\% of gelatine and 3\% of agar. In this kind of medium and frequency range, the single scattering contribution is by far predominant \cite{aubry}. The thickness $L$ of the scattering slab is 100 mm. Once the inter-element matrix $\mathbf{H}$ is measured, the short-time Fourier analysis described in Sec.\ref{sec:exp_num} yields a set of response matrices $\mathbf{K}(T,f)$. Then, the separation of single and multiple scattering is achieved as described in Sec.\ref{sec:separation_ss_ms}. 

Fig.\ref{fig:fig5} shows a typical experimental result, taken at time $T = 114$ $\mu$s and frequency $f = 3.05$ MHz. Note that the separation rank between the signal and noise subspaces is here $p = 2$, which confirms that Eq.\ref{eqn:eq_ss_signal_3} doesnot strictly hold.. $\mathbf{K^S}$ exhibits the deterministic coherence along the antidiagonals (Fig.\ref{fig:fig5}(a)) which is characteristic of single scattering. Obviously, $\mathbf{K^S}$ is very close to the raw matrix $\mathbf{K}$ (Fig.\ref{fig:fig2}), since single scattering is predominant. As to $\mathbf{K^M}$, it displays a random feature as expected for the multiple scattering contribution (Fig.\ref{fig:fig5}(b)). However, one cannot conclude that it originates in multiple scattered waves: it could also correspond to experimental noise.
\begin{figure}[htbp] 
\includegraphics{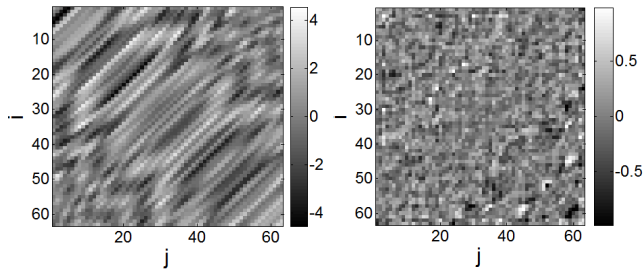}
\caption{\label{fig:fig5}Separation of single and multiple scattering contributions at time $T=114$ $\mu$s and frequency $f=3.05$ MHz. (a) Real part of $\mathbf{K^S}$. (b) Real part of $\mathbf{K^M}$.}
\end{figure}

In order to establish the multiple-scattering origin of $\mathbf{K^M}$, we calculated the mean backscattered intensity $I^M$ as a function of the source-receiver distance $X=x_j-x_i=m \delta x$ and the arrival time $T$ 
\begin{equation}
\label{eqn:mean_intensity}
I^M(X,T)=\left < \left | \tilde{k}^M_{ij}(T,f) \right |^2 \right >_{f,\{(i,j)\,|\, m=j-i\}}
\end{equation}
The symbol $\left < . \right>$ denotes an average over the quantities in the subscript, here frequency and all source/receiver couples $(i,j)$ separated by the same distance $X$. In Fig.\ref{fig:fig6}, the spatial dependence of $I^M$ is compared with the total intensity, at a given time $T$. Whereas the total intensity $I$ shows no preferred direction, $I^M(X)$ exhibits a typical signature of multiple scattering: the coherent backscattering peak clearly arises around $X=0$.
\begin{figure}[htbp] 
\includegraphics{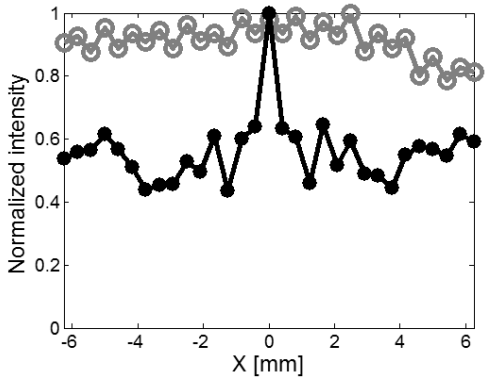}
\caption{\label{fig:fig6}The multiple scattering intensity $I^M$ (black dots) and the total intensity $I$ (grey circles) are plotted versus $X$, at time $T=137$ $\mu$s. The intensity profiles have been renormalized with their maximum.}
\end{figure}
This phenomenon has been widely observed and studied in wave physics (optics \cite{Ishimaru1,wolf,albada2,akkermans2,akkermans3,wiersma,labeyrie1,labeyrie2}, acoustics \cite{bayer,tourin2,yamamoto,rosny,rosny3}, seismology \cite{larose,tiggelen,margerin}). It manifests itself as an enhancement, by a factor 2, in the backscattered intensity at the vicinity of the source (\textit{i.e} $X=0$). Its physical origin lies in the constructive wave interference between reciprocal paths that have been scattered at least twice; it can only appear when multiple scattering occurs and the reciprocity symmetry is preserved. The intensity profile shown in Fig.\ref{fig:fig6} is a spectacular evidence of multiple scattering and shows the efficiency of our technique for extracting the multiple-scattering waves among a predominant single scattering contribution.

Interestingly, though it is weak, the multiple scattering contribution can be taken advantage of in order to characterize the medium and determine separately the scattering losses and the absorption losses. When a wave propagates through a random medium, it loses progressively its coherence: after traveling over a distance $L$, only a fraction $\exp \left ( -  {L}/{l_{ext}} \right)$ of the initial energy still propagates in coherence with the initial wave. The parameter $l_{ext}$, called the extinction mean-free path, characterizes the extinction length of the coherent part of the wave. It comes from two distinct phenomena (scattering and intrinsic absorption of the medium) which are associated to two characteristic lengths: the elastic
mean-free path $l_e$ and the absorption mean-free path $l_a$, such that
\begin{equation}
\label{eqn:mean_free_path}
\exp \left ( -  {L}/{l_{ext}} \right)= \exp \left ( - {L}/{l_{e}} \right) \times \exp \left ( -  {L}/{l_{a}} \right)
\end{equation}
Experimentally, $l_{ext}$ can be determined by measurements of the ensemble-averaged field transmitted through a scattering layer \cite{page,zhang,scales1,scales2,derode4}. However, this kind of experiments does not allow to distinguish $l_a$ from $l_e$. 

\begin{figure}[htbp] 
\includegraphics{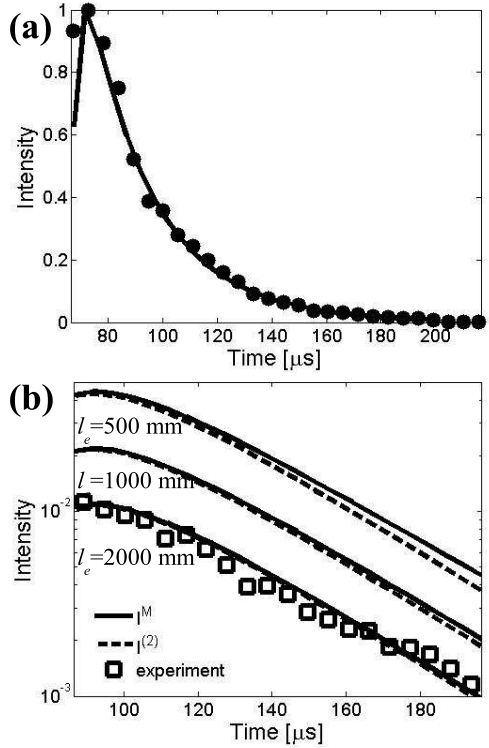}
\caption{\label{fig:fig7}(a) Single-scattered intensity $I^S(X=0,T)$ versus time. Experimental
measurements (black circles) are fitted with the theoretical curve (continuous black line) considering an extinction length $l_{ext} = 50$ mm. (b) Multiply-scattered intensity $I^M(X=0,T)$ versus time. Experimental measurements (white squares) are compared with thoretical results for $I^M(X=0,T)$(continuous black lines) and $I^{(2)}(X=0,T)$(dashed black lines) for different values of the mean free path, while keeping $l_{ext} = 50$ mm. All intensities have been normalized by the maximum of the single scattered intensity $I^S(X=0,T)$ over time.}
\end{figure}
We focus on the single and multiple scattering intensities obtained at the source: $I^S(0,T)$ and $I^M(0,T)$. They are plotted in Fig.\ref{fig:fig7}. Note that the intensity of the multiple scattering contribution is less than 1\% of the single scattering contribution. Once $I^S(0,T)$ and $I^M(0,T)$ have been measured, we can fit both experimental curves with $l_a$ and $l_e$ as independent adjustable parameters. To that end, we need a theoretical model describing the spatial and temporal evolution of the mean intensity inside the random medium. In the literature, the mean intensity is often assumed to obey the diffusion equation \cite{akkermans}. The diffusion approximation is simple, but only valid in the long-time limit. Since we deal with a weakly scattering medium, the elastic mean free path is expected to be very large compared to the scattering path lengths ($l_e >> cT$). Thus, the diffusion approximation does not apply to our problem. Instead, we used the radiative transfer equation (RTE) \cite{rossum}. Paasschens \cite{paasschens} proposed an exact solution of the RTE in time-domain and real space for an infinite 2D random medium (see Appendix C), Eq.C21). Based on this theoretical work, we have computed the exact expression of the single and double scattering intensities, $I^S(0,T)$ and $I^{(2)}(0,T)$, as well as an approximate expression of the multiple scattering intensity $I^M(0,T)$, considering the medium as semi-infinite and two-dimensional. 

The choice of a 2D model is justified as follows. Experimentally, the transducers are 10 mm in height, which is much larger than the average wavelength (0.5 mm). Moreover, a vertical cylindrical acoustic lens ensures that the emitted beam remains collimated in the $(x,z)$ plane. Similarly, in reception only waves propagating in the $(x,z)$ plane are recorded by the transducers. Thus the single scattering problem is clearly 2D. As to multiple scattering, for the same reason the only paths that can generate a signal on the receiving transducers are those whose first and last scatterer are in the $(x,z)$ plane. The gel sample being weakly scattering, $I^M$ is mostly dominated by double scattering (see Fig.\ref{fig:fig7}(b)). Thus even though the wave propagation in the gel sample is 3D, we have used the 2D solution for the RTE.

The detailed calculations of $I^S$, $I^{(2)}$ and $I^M$ are shown in Appendix C. It appears that the single scattering intensity $I^S(0,T)$ exhibits a temporal evolution which only depends on the extinction length $l_{ext}$. In the case of the gel studied here, the best fit of the experimental results yields $l_{ext}= 50$ mm (Fig.\ref{fig:fig7}(a)). Once $l_{ext}$ is known, $l_e$ and $l_a$ can be determined by fitting $I^M(0,T)$ with theory (Fig.\ref{fig:fig7}(b)) with only one adjustable parameter since $1/l_{ext}=1/l_e+1/l_a$. The scattering gel is found to be much more absorbing than scattering : $l_e \sim 2000$ mm, while $l_a \sim 50$ mm. Fig.\ref{fig:fig7}(b) also displays the theoretical evolution of the double scattering contribution $I^{(2)}(0,T)$. For $l_e \sim 2000$ mm, $I^{(2)}$ and $I^M$ are nearly identical, which shows that the double scattering contribution clearly dominates the multiple scattering intensity in the gel sample. As the theoretical expression derived in Appendix C is exact for $I^{(2)}$, the measured values of $l_e$ and $l_a$ are reliable.

{In this example, the medium was a weakly scattering gel, with a ratio $I^M/I^S$ less than 1\%. Yet the separation of single and multiple scattering contributions can also be achieved in real scattering media for which $I^M/I^S$ is closer to unity, as we present in the next section.}

\section{\label{sec:breast}Application to human soft tissues}

\begin{figure}[htbp] 
\includegraphics{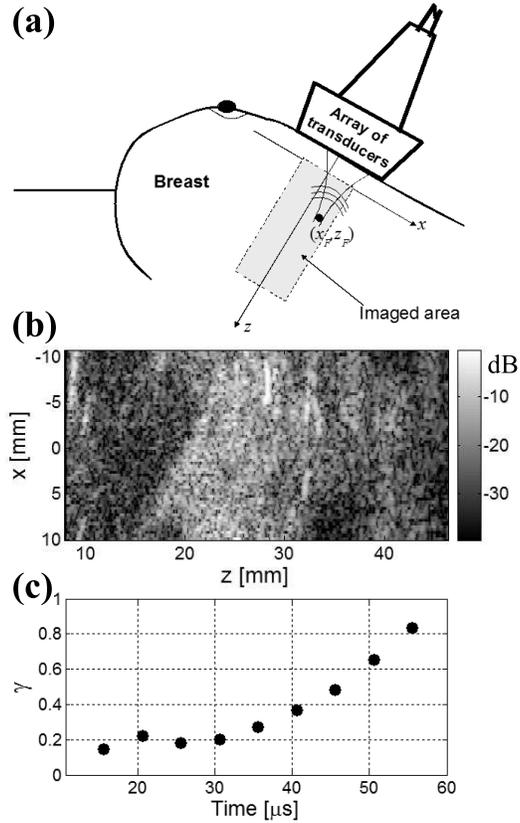}
\caption{\label{fig:fig8}(a) Experimental setup used for the investigation of multiple scattering in breast tissues. (b) Echographic image of the breast. The grey scale is in dB. (c) Multiple scattering rate $\gamma$ as a function of time.}
\end{figure}
The same kind of experiment has been performed in a biological medium for which ultrasound is often used: the breast. The experimental set up is depicted in Fig.\ref{fig:fig8}(a). We use a $N$-element ultrasonic array ($N = 125$) with a 4.3 MHz central frequency and a 3.5-5MHz bandwidth; the array pitch $\delta x$ is 0.33 mm. The emitted signal is a 0.7-$\mu$s sinusoidal burst at $f_c = 4.3$ MHz. The sampling frequency is 50 MHz. The experimental procedure is the same as in Sec.\ref{sec:exp_num}. The separation of single and multiple scattering contributions can be performed as in Sec.\ref{sec:separation_ss_ms}, but an adjustment has to be made. Indeed, in our experimental configuration, the array of transducers is placed in the \textit{near-field} of the scattering medium ($a=0$). Consequently, the entries of matrix $\mathbf{K}$ are not identically distributed: the variance (i.e the mean intensity $I$) of $k_{ij}$ decreases significantly with the distance $X=x_j-x_i$ between the source and the receiver, as shown by Fig.\ref{fig:fig9}. This implies a different variance for each line of matrix $\mathbf{A}$, hence modifying its theoretical distribution of singular values. The upper bound $\lambda_{\mbox{\small{max}}}^{(0)}$ can be computed numerically taking into account this non-uniform distribution of matrix elements (See Appendix D). However this is only possible at the first iteration ($q=1$). Indeed, whereas the variance of $\mathbf{A}$ can be estimated, the variance distribution of the subspaces of $\mathbf{A}$ is unknown a priori. Unless we do the (strong) approximation that this variance is uniform, in which case Eq.\ref{eqn:dens_sing_val_3_chap3} could be applied, {we cannot follow the procedure described in Sec.\ref{sec:separation_ss_ms}}. Here, since $\mathbf{A}$ clearly has a non-uniform variance, by precaution we choose a different strategy: the same upper bound $\lambda_{\mbox{\small{max}}}^{(0)}$ is considered at each iteration $q$ of the single/multiple scattering separation process (see Sec.\ref{sec:separation_ss_ms}). Since $\lambda_{\mbox{\small{max}}}^{(q)} < \lambda_{\mbox{\small{max}}}^{(0)}$, this precaution tends to overestimate the threshold {and decrease the probability of error.}

Fig.\ref{fig:fig9} compares the multiple scattering, single scattering and total intensity profiles, at a given time $T$. Contrary to the previous experiment in the gel sample (see Fig.\ref{fig:fig6}), the spatial intensity profiles, $I(X)$ and $I^S(X)$, are not flat due to the \textit{near-field} configuration of the experiment. Once again, $I^M(X)$ exhibits a coherent backscattering peak on top of a flat incoherent intensity with an enhancement factor close to 2. Interestingly, this indicates that $\mathbf{K^M}$ is not a noise contribution, but does originate from multiple wave scattering in the breast tissue, even though we operate at a frequency (4.3MHz) for which human soft tissues are usually treated as single-scattering.
\begin{figure}[htbp] 
\includegraphics{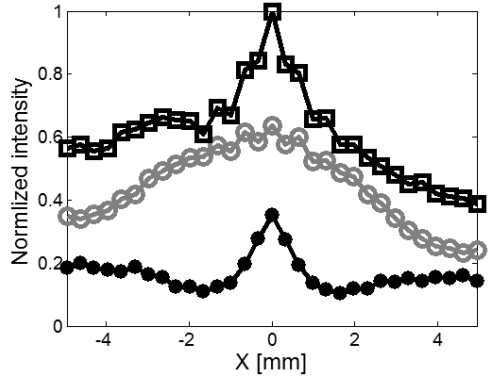}
\caption{\label{fig:fig9}The multiple scattering intensity $I^M(X)$ (black dots), the single scattering intensity $I^S(X)$ (grey circles) and the total intensity $I(X)$ (black squares) are plotted versus $X$, at time $T=35.6$ $\mu$s. The intensity profiles have been renormalized with the maximum of the total intensity.}
\end{figure}

An ultrasound image of the breast has also been obtained with the same array, using 63-element subapertures (Fig.\ref{fig:fig8}(b)). As usual in ultrasound imaging, focusing in emission and reception is achieved by applying a set of 63 time delays to the signals transmitted/received by the array. The time delays are computed in order to focus at the desired region of interest, centred at coordinates $z_F$ and $x_F$, assuming that the velocity of sound in soft tissues is known. In the case of breast, as for most soft biological tissues, it is close to that of water ($c=1500$ m/s). Here, 2666 focal planes ($z_F =$ 8 mm to 48 mm) and 63 values of $x$ ($x_F=-31 \delta x$ to $x_F=31 \delta x$ with $\delta x$ the array pitch) have been used.  At each depth $z_F$, a new set of time-delays is calculated; this is more demanding that classical ultrasound imaging techniques, which generally use the same set of time-delays as long as $z_F$ is within the depth of field. A line of the resulting image represents, in gray level, the amplitude of the total echographic signal at the focal time $T=2z_F/c$, once focused beamforming has been applied to the 63 received signals. The resulting image displays the reflectivity of the medium under investigation. Ultrasound images of human tissues usually reveal the interfaces of inner organs, and often exhibit a speckled appearance due do random scattering by sub-wavelength inhomogeneities (cells, fibers, tissues \textit{etc.}) \cite{Burckhardt}. Here the scanned area is particularly echogene between 30-40 $\mu$s, corresponding to the depth range 22.5-30 mm. The typical ultrasound image of a human organ (here, the breast) is representative of the amplitude of backscatter at a given time which hopefully (under the single scattering assumption) corresponds to a given depth, but it does not allow us to distinguish single and multiple scattering contributions.

However, once the matrix $\mathbf{K}$ has been recorded, not only can we build an echographic image but we can also isolate the multiple scattering contribution, {and estimate the multiple scattering rate $\gamma$ (Eq.\ref{G}). $\gamma$ has been averaged over the whole frequency spectrum  and is displayed on Fig.\ref{fig:fig8}(c) as a function of time.} Fig.\ref{fig:fig8}(c) complements the information brought by the echographic image. A relevant observation is that multiple scattering becomes predominant from $T = 46$ $\mu$s, that is to say beyond a depth of 34.5 mm. It means that the single scattering assumption, upon which the imaging process is based, is incorrect. It does not mean however that the image is totally wrong; a rate $\gamma$ of $50 \%$ means that half of the intensity received by one individual array element comes from multiple scattering. In classical array imaging, each line of the picture is constructed by focused beamforming {in emission and reception}. This procedure reduces the importance of multiple scattering in the final image because the single scattered contributions coming from a target in the focal zone add up coherently whereas the contributions from multiple scattering can be expected to be uncorrelated. With $\gamma=0.5$ and assuming $N'=63$ uncorrelated array elements, the multiple scattering rate becomes $\sqrt{\gamma/(1-\gamma)}/{N'} \sim 1/63$ after beamforming. This is probably an underestimation since multiple scattering signals cannot be fully uncorrelated \cite{aubry5,aubry3}. As a result, the proportion of multiple scattering in the final image around $T = 50$ $\mu$s is of the order of a few percents, which is still weak. Also note that at larger times, the technique we presented here would fail: after $T = 50$ $\mu$s the rate of multiple scattering becomes too large for the SVD to extract the single scattering contribution ($p = 0$), at least for some frequencies of the spectrum. As the results are averaged over the whole frequency spectrum (3.5-5 MHz), the multiple scattering rates that are presented here are still meaningful, until $T=60$ $\mu$s. Beyond that time, the method we presented here would be inadequate to separate single and multiple scattering.

\section{\label{sec:conclusion}Conclusion}
The approach we developed here can separate single- and multiple-scattered waves in randomly heterogeneous media. It requires an array of transmitters/receivers and takes advantage of the persistence of a deterministic coherence of single scattering signals along the antidiagonals of the inter-element matrix. Once a singular value decomposition is applied, the single scattering contribution (signal subspace) is separated from the multiple scattering contribution (noise subspace) by using a criterion based on random matrix theory. Unlike previous works \cite{aubry5,aubry4} this technique is particularly well-suited for weakly scattering media, for which single scattering dominates. In such media, the technique we presented here is not intended to enhance the quality of the ultrasound image, but rather to complement it, in two ways. Firstly, the experimental results indicate that this approach can be applied for characterization purposes: the separation of single and multiple scattering provides a way to measure the scattering and absorption mean-free paths independently. This idea was tested on a synthetic gel. Secondly, the technique was also applied in vivo to the case of breast imaging with ultrasonic waves around 4.3 MHz. The occurrence of multiple scattering has been established and its contribution to the backscattered wave-field is shown to be far from negligible. By measuring the relative amount of multiple scattering, the method serves as an experimental test for the first Born approximation (single scattering), which is usually made in such tissues.

\begin{acknowledgments}
The authors would like to thank Pr Mickael Tanter for fruitful discussions and Patricia Daenens for her technical help, as well as the \textit{groupe de recherches} IMCODE of CNRS (GDR 2253).
\end{acknowledgments}

\appendix

\section{\label{app:gain}}
Single and multiple scattering signals are not strictly orthogonal. Thus, $\mathbf{S}$ can be expressed as the sum of the single scattering contribution $\mathbf{A^S}$ plus a residual multiple scattering noise $\mathbf{R}$:
\begin{equation}
\label{eqn:appB_03}
\mathbf{S}=\mathbf{A^S}+\mathbf{R}
\end{equation}
The aim of this appendix is to estimate the remaining part of multiple scattering which corrupts the signal subspace $\mathbf{S}$. In other words, we want to determine the mean intensity $\sigma_R^2$ of coefficients $r_{lm}$ ($\sigma_R^2= \left < \left | r_{lm} \right |^2 \right >$). 

To that aim, we apply the first order perturbation theory \cite{cohen} to the autocorrelation matrix $\mathbf{AA^{\dag}}$. Actually, $\mathbf{AA^{\dag}}$ can be decomposed as a sum of an unperturbed autocorrelation matrix $\mathbf{A^SA^{S\dag}}$ (linked with single scattering) and a perturbation matrix $\mathbf{W}$ due to multiple scattering :
\begin{gather}
\mathbf{AA^{\dag}}=\mathbf{A^SA^{S\dag}}+\mathbf{W} \mbox{,} 
\label{eqn:appB_1}
\\
\mbox{with }\mathbf{W}=\underbrace{\mathbf{A^SA^{M\dag}}+\mathbf{A^MA^{S\dag}}}_{1^{\mbox{\small st}}\mbox{ order}}+\underbrace{\mathbf{A^MA^{M\dag}}}_{2^{\mbox{\small nd}} \mbox{ order}}
\label{eqn:appB_2}
\end{gather}
As we deal with weakly scattering media, the single scattering contribution is predominant: $\sigma_S >> \sigma_M$. Thus, $\mathbf{W}$ can be seen as a low perturbation of the autocorrelation matrix. Moreover, we can neglect the second order term $\mathbf{A^MA^{M\dag}}$ in Eq.\ref{eqn:appB_2}. $\mathbf{W}$ can be simplified into 
\begin{equation}
\label{eqn:appB_3}
\mathbf{W}\simeq \mathbf{A^SA^{M\dag}}+\mathbf{A^MA^{S\dag}}
\end{equation}

Let us first focus on the reference state, \textit{i.e} with no multiple scattering. In this case, we have
\begin{equation}
\label{eqn:appB_0}
\mathbf{A}\equiv \mathbf{A^S}=\sum_{k=1}^p\lambda_{k}^{o}\mathbf{U^{(o)}_{k}}\mathbf{V^{(o)\dag}_{k}}\mbox{ , for }\sigma_M \equiv 0
\end{equation} 
where $\lambda^o_k$ denotes the $k^{\mbox{\small th}}$ singular value of $\mathbf{A^S}$. $\mathbf{U_k^o}$ and $\mathbf{V_k^o}$ are the singular vectors associated to $\lambda_{k}^o$. 

The perturbation theory can now be applied to $\mathbf{AA^{\dag}}$. At first order, this theory states that only the eigenvalues $\left ( \lambda_k\right) ^2$ of $\mathbf{AA^{\dag}}$ are perturbed by multiple scattering. The singular vectors remain identical to those obtained in the unperturbed case (\textit{i.e.} without multiple scattering), hence $\mathbf{U_k} \simeq \mathbf{U_k^o}$ and $\mathbf{V_k} \simeq \mathbf{V_k^o}$. The $p$ first eigenvalues $\left (\lambda_k \right )^2$ of $\mathbf{AA^{\dag}}$ can be written as \cite{cohen}:
\begin{equation}
\label{eqn:appB_4}
\left (\lambda_k \right )^2 \simeq \left (\lambda_k^o \right )^2+\mathbf{U_k^{o\dag} W U_k^o}
\end{equation} 
$\left (\lambda_k^o \right )^2$ is the unperturbed eigenvalue (that we would obtain without multiple scattering) and the term $\mathbf{U_k^{o\dag} W U_k^o}$ corresponds to the perturbation due to multiple scattering. 

Using Eqs.\ref{eqn:appB_4} \& \ref{eqn:appB_3}, one can try to express $\left (\lambda_k \right )^2-\left (\lambda_k^o \right )^2$:
\begin{eqnarray}
\left ( \lambda_k\right) ^2-\left ( \lambda^o_k\right) ^2 & \simeq & \mathbf{U_k^{o\dag} W U_k^o} \nonumber\\
& \simeq & \mathbf{U_k^{o\dag} A^S A^{M\dag} U_k^o} + \mathbf{U_k^{o\dag} A^M A^{S\dag} U_k^o}  \nonumber\\
 & \simeq & \lambda_k^o \mathbf{V_k^{o\dag} A^{M\dag} U_k^o} +  \lambda_k^o   \mathbf{U_k^{o\dag} A^M V_k^o} \nonumber\\
\left ( \lambda_k\right) ^2-\left ( \lambda^o_k\right) ^2 & \simeq & 2 \lambda_k^o \Re \left \{   \mathbf{U_k^{o\dag} A^M V_k^o} \right\}
\label{eqn:appB_5}
\end{eqnarray}
$\mathbf{U_k^{o\dag} A^M V_k^o} $ is a random complex variable with zero mean. Let us calculate its variance:
\begin{eqnarray}
 \mbox{var} \left [   \mathbf{U_k^{o\dag} A^M V_k^o} \right] & = & \left < \mathbf{U_k^{o\dag} A^M V_k^oV_k^{o\dag} A^{M\dag} U_k^o} \right> \nonumber\\
 & = & \left <  \mathbf{U_k^{o\dag}} \mathbf{ A^M  A^{M\dag} } \mathbf{U_k^o} \right > \nonumber\\
  & = & \mathbf{U_k^{o\dag}} \left < \mathbf{ A^M  A^{M\dag} }\right  > \mathbf{U_k^o}
\label{eqn:appB_6}
\end{eqnarray}
We assume that the matrix $\mathbf{A^M }$ is random, hence the ensemble average of $ \mathbf{A^M  A^{M\dag}}$ yields 
\begin{equation}
\label{eqn:appB_7}
\left < \mathbf{ A^M  A^{M\dag} }\right  > = L\sigma_M^2 \mathbf{I}
\end{equation}
where $\mathbf{I}$ is the identity matrix and $\sigma_M^2$ is the mean power of multiple scattering signals: $\left < \left | a^M_{lm} \right |^2 \right >=\sigma_M^2$. The combination of Eqs.\ref{eqn:appB_6} \& \ref{eqn:appB_7} yields:
\begin{equation}
\label{eqn:appB_8}
\mbox{var} \left [   \mathbf{U_k^{o\dag} A^M V_k^o} \right]= L\sigma_M^2 
\end{equation}

Let us calculate the variance of $\left ( \lambda_k\right) ^2-\left ( \lambda^o_k\right) ^2$ (Eq.\ref{eqn:appB_5}):
\begin{eqnarray}
\left < \left [ \left ( \lambda_k\right) ^2-\left ( \lambda^o_k\right) ^2 \right]^2 \right>& \simeq & 4 \left ( \lambda^o_k\right) ^2 \mbox{var} \left [ \Re \left \{   \mathbf{U_k^{o\dag} A^M V_k^o} \right \} \right] \nonumber \\
& \simeq & 4 \left ( \lambda^o_k\right) ^2 \frac{\mbox{var} \left [    \mathbf{U_k^{o\dag} A^M V_k^o}  \right]}{2}
\label{eqn:appB_9}
\end{eqnarray}
By injecting Eq.\ref{eqn:appB_8} in Eq.\ref{eqn:appB_9}, we obtain
 \begin{equation}
\label{eqn:appB_10}
 \left < \left [ \left ( \lambda_k\right) ^2-\left ( \lambda^o_k\right) ^2 \right]^2 \right> \simeq 2 \left ( \lambda^o_k\right) ^2 L\sigma_M^2  
\end{equation}
Using the approximation $\left ( \lambda_k\right) ^2 - \left ( \lambda^o_k\right) ^2 \simeq 2 \lambda^o_k \left ( \lambda_k- \lambda^o_k \right)$, the last equation becomes:
 \begin{equation}
\label{eqn:appB_11}
 \left <   \epsilon_k^2 \right> \simeq \frac{ L\sigma_M^2 }{2} 
\end{equation}
where $ \epsilon_k=\lambda_k -\lambda_k^o$.

Using Eq.15 of the manuscript and Eq.\ref{eqn:appB_0} and the fact that $\mathbf{U_k} \simeq \mathbf{U_k^o}$ and $\mathbf{V_k} \simeq \mathbf{V_k^o}$, the matrix $\mathbf{R}=\mathbf{S-A^S}$ can be expressed as a function of $ \epsilon_k$:
 \begin{equation}
\label{eqn:appB_13}
\mathbf{R} \simeq \sum_{k=1}^p \epsilon_k \mathbf{U_k^o}\mathbf{V^{o\dag}_{k}} 
\end{equation}
Now, we can estimate the variance $\sigma_R^2$ of the coefficients $r_{lm}$:
 \begin{eqnarray}
\sigma_R^2 &= & \left < \left | r_{lm} \right |^2 \right> \nonumber \\
&\simeq & \sum_{k=1}^p \left < \left | \epsilon_k \right |^2 \right> \left < \left | u^o_{kl} \right |^2 \right> \left < \left |v^o_{km} \right |^2 \right> 
\label{eqn:appB_14}
\end{eqnarray}
Because the singular vectors $\mathbf{U_k^o}$ and $\mathbf{V_k^o} $ are normalized, $\left < \left | u^o_{kl} \right |^2 \right>=1/{M} $ and $\left < \left |v^o_{km} \right |^2 \right> =1/{L} $. Using Eq.\ref{eqn:appB_11}, we finally obtain:
 \begin{equation}
\label{eqn:appB_15}
\sigma_R^2 \simeq \frac{p}{2M} \sigma_M^2
\end{equation}
The mean power of the residual multiple scattering noise is $\sigma_R^2 \simeq p \sigma_M^2/(2M)$. This residual noise is negligible compared to single scattering signals since $\sigma_S^2>>\sigma_M^2$.

\section{\label{app:app3}}

This appendix describes the numerical process we use to determine the upper bound $\lambda_{\mbox{\small{max}}}^{(q)}$ of the singular values distribution, when short-range correlations between matrix entries exist. 

The first step is to estimate these correlations. This is done by measuring the correlation coefficients $\Gamma_m^E$ and $\Gamma_m^R$ between respectively the lines and columns of $\mathbf{A}$:
\begin{eqnarray}
\Gamma^E_m & = & \frac{\left < a_{i,j}a_{i+m,j}^*\right >_{(i,j)}}{\left < \left | a_{i,j}\right |^2 \right >_{(i,j)}} \\
\Gamma^R_m& = & \frac{\left < a_{i,j}a_{i,j+m}^*\right >_{(i,j)}}{\left < \left | a_{i,j}\right |^2 \right >_{(i,j)}}
\end{eqnarray}
where the symbol $<. >$ denotes an average over the variables in the subscript, \textit{i.e} the source/receiver pairs $(i,j)$. The integer $m$ represents the distance between sources and receivers, in units of the array pitch $\delta x$.

The second step consists in building two correlation matrices $\mathbf{C}$ (of size $(M-q) \times (M-q)$) and $\mathbf{D}$ (of size $L \times L$) from the measured correlation coefficients $\Gamma^E_m$ and $\Gamma^R_m$. The coefficients $c_{ij}$ and $d_{ij}$ of these correlation matrices are given by:
\begin{eqnarray}
c_{ij} & = & \Gamma^E_{i-j} \\
d_{ij}& = & \Gamma^R_{i-j}
\end{eqnarray}
Once these correlation matrices are built, the next step consists in generating a random matrix which displays the same short-range correlations as matrix $\mathbf{A}$. 

To that aim, a random matrix $\mathbf{P}$ of dimension $(M-q) \times L$ is first generated numerically. Its coefficients are totally independent and identically distributed. Then, a matrix $\mathbf{Q}$ is built from $\mathbf{P}$, such that
\begin{equation}
\mathbf{Q}=\mathbf{C}^{\frac{1}{2}}\mathbf{P}\mathbf{D}^{\frac{1}{2}} 
\end{equation}
The matrix $\mathbf{Q}$ exhibits the same correlation properties as $\mathbf{A}$. The SVD of $\mathbf{Q}$ is then achieved and its first singular value $\lambda_1$ is renormalized according to
\begin{equation}
\tilde{\lambda}_1=\frac{\lambda_1}{\sqrt{\frac{1}{M-q}\sum_{k=1}^{M-q}\lambda^{2}_k}}
\end{equation}
By repeating this operation over 500 realizations, we can build a cumulative histogram of the first singular value $\tilde{\lambda}_1$. The distribution function $F_1$ of the first singular value, $F_1(\lambda)=\mbox{Prob} \{ \tilde{\lambda}_1 < \lambda\}$, can be estimated. The upper bound $\lambda_{\mbox{\small{max}}}^{(q)}$ is then deduced from $F_1$. An acceptable probability of error $\gamma$ is first set (typically, $\gamma=10^{-2}$), hence the upper bound $\lambda_{\mbox{\small{max}}}^{(q)}$:
\begin{equation}
\label{eqn:threshold}
\lambda_{\mbox{\small{max}}}^{(q)}=F^{-1}_1(1-\gamma)
\end{equation}

\section{\label{app:intensity_calcul}}
In this appendix, we derive an expression for the single and multiple-scattering contributions to the average backscattered intensity, based on radiative transfer theory. Radiative transfer theory describes the spatial and temporal dependence of the radiance (or specific intensity) $P(\mathbf{r},t,\mathbf{u})$ in a random medium. Radiance is defined as energy flow, propagating in the direction $\mathbf{u}$, per unit normal area per unit solid angle $d \Omega$ per unit time. It follows a transport (Boltzmann) equation :
\begin{equation}
\label{radiative}
\frac{1}{c}\frac{\partial}{ \partial t}P(\mathbf{r},t,\mathbf{u})+\mathbf{u}. \nabla P(\mathbf{r},t,\mathbf{u}) + l_{ext}^{-1}P(\mathbf{r},t,\mathbf{u}) = l_e^{-1}P(\mathbf{r},t)+ c^{-1}S(\mathbf{r},t,\mathbf{u}),
\end{equation}
with $S$ the source term {and $P(\mathbf{r},t)$ the intensity, defined as the angular average of the radiance}
\begin{equation}
P(\mathbf{r},t)=\frac{1}{2 \pi}\int d \Omega P(\mathbf{r},t,\mathbf{u})
\end{equation}

Classically, the transport equation can be derived from the Bethe-Salpether equation \cite{frisch,margerin2}, neglecting all interference (coherent) effects. Here we adapt the theoretical developments of Paasschens \cite{paasschens}, which were derived for an infinite random medium, to our experimental configuration. The problem is solved in two dimensions.

The typical configuration is depicted in Fig.1 of the manuscript. The random medium $(\vartheta)$ is assumed to be semi-infinite, with a plane interface. It is characterized by an elastic mean-free path $l_e$ and an absorption length $l_a$. The scattering events in the random medium are assumed to be isotropic, hence there is no difference between the elastic mean-free path and the transport mean-free path $l^*$.

In our case, the medium is not statistically invariant under translation since it is semi-infinite: $P$ depends both on the observer ($\mathbf{r}$) and the source ($\mathbf{r_s}$). The radiative transfer equation is modified into:
\begin{eqnarray}
\frac{1}{c}\frac{\partial}{ \partial t}P(\mathbf{r_s},\mathbf{r},t,\mathbf{u})+\mathbf{u}. \nabla P(\mathbf{r_s},\mathbf{r},t,\mathbf{u}) &+& M(\mathbf{r}) l_{ext}^{-1}P(\mathbf{r_s},\mathbf{r},t,\mathbf{u}) \nonumber \\
&=&  M(\mathbf{r}) l_e^{-1}P(\mathbf{r_s},\mathbf{r},t)+ c^{-1}\delta(t) \delta(\mathbf{r}-\mathbf{r_s}),
\end{eqnarray}
{where an isotropic point source is now considered: $S(\mathbf{r},t,\mathbf{u})=\delta(t) \delta(\mathbf{r}-\mathbf{r_s})$.} $M(\mathbf{r})$ accounts for the semi-infinite nature size of the random medium:
\begin{equation}
M(\mathbf{r'})= 
\left \{ 
\begin{array}{l}
1 \mbox{, if }\mathbf{r'} \in (\vartheta)\\
0 \mbox{, otherwise}
\end{array} 
\right.
\label{eqn:M}
\end{equation}

This problem can be solved by considering separately the contributions to intensity from $N=0,1,2,\cdots $ scattering events
\begin{equation}
P(\mathbf{r_s},\mathbf{r},t,\mathbf{u})=\sum_{N=0}^{\infty} P_N(\mathbf{r_s},\mathbf{r},t,\mathbf{u})\mbox{,    } 
P(\mathbf{r_s},\mathbf{r},t)=\sum_{N=0}^{\infty} P_N(\mathbf{r_s},\mathbf{r},t)
\end{equation}
The partial intensities $P_N$ satisfy
\begin{eqnarray}
\label{eqn:ballistic_eq}
\left ( \frac{\partial}{c \partial t} + \mathbf{u}.\nabla + M(\mathbf{r}) l_{ext}^{-1} \right) P_0(\mathbf{r_s},\mathbf{r},t,\mathbf{u}) & =& c^{-1} \delta(t) \delta(\mathbf{r}) \\
\label{eqn:partial_eq}
\left ( \frac{\partial}{c \partial t} + \mathbf{u}.\nabla + M(\mathbf{r}) l_{ext}^{-1} \right) P_N(\mathbf{r_s},\mathbf{r},t,\mathbf{u}) &= & M(\mathbf{r}) l_e^{-1} P_{N-1}(\mathbf{r_s},\mathbf{r},t)\mbox{, for }N>0
\end{eqnarray}

Let us first investigate the ballistic intensity $P_0$. Following Paaschens \cite{paasschens}, the differential operators on the left hand side of Eq.\ref{eqn:ballistic_eq} can be integrated, to yield
\begin{eqnarray}
P_0(\mathbf{r_s},\mathbf{r},t,\mathbf{u}) &= &\frac{1}{c}\int_0^{\infty} dr_0 e^{-f(\mathbf{r_s},\mathbf{r_s}+r_0\mathbf{u})/l_{ext}} \delta(\mathbf{r}-r_0 \mathbf{u}-\mathbf{r_s}) \delta(t-r_0/c) \nonumber \\
&=& e^{-f(\mathbf{r_s},\mathbf{r_s}+ct\mathbf{u} )/l_{ext}} \delta(\mathbf{r}-ct\mathbf{u} -\mathbf{r_s}) \nonumber \\
&=&\frac{e^{-f( \mathbf{r_s},\mathbf{r})/l_{ext}}}{|\mathbf{r}-\mathbf{r_s}|} \delta(|\mathbf{r}-\mathbf{r_s}|-ct) \delta \left ( \mathbf{u}-\mathbf{u_{s,r}} \right) \label{eqn:ballistic_radiance} \\
\mbox{where  }f(\mathbf{r_s},\mathbf{r}) & = &\int_0^{|\mathbf{r}-\mathbf{r_s}|} \! \! \! \!  \! \! \! \! dr' M \left ( \mathbf{r_s}+r'\mathbf{u_{s,r}}\right)
 \label{eqn:def_f} \\
 \mbox{and  }\mathbf{u_{s,r}}&=&\frac{\mathbf{r}-\mathbf{r_s}}{|\mathbf{r}-\mathbf{r_s}|}
\nonumber
\end{eqnarray}
The Dirac distribution $\delta(|\mathbf{r}-\mathbf{r_s}|-ct)$ describes the propagation of the energy pulse at the finite speed $c$. $|\mathbf{r}-\mathbf{r_s}|^{-1}$ accounts for the geometrical spreading of the ballistic radiance in two dimensions. $f(\mathbf{r_s},\mathbf{r})$ is the ballistic path length within the scattering medium:
\begin{itemize}
\item If $\mathbf{r_s},\mathbf{r}\in (\vartheta)$, then $f(\mathbf{r_s},\mathbf{r})=|\mathbf{r}-\mathbf{r_s}|$.
\item If $\mathbf{r_s},\mathbf{r}\notin (\vartheta)$, then $f(\mathbf{r_s},\mathbf{r})=0$.
\item In other configurations (see Fig.\ref{fig:fig11}), the ballistic radiance propagates partly in free space, partly in the random medium: $0<f(\mathbf{r_s},\mathbf{r})<|\mathbf{r}-\mathbf{r_s}|$.
\end{itemize}
The angular average of  Eq.\ref{eqn:ballistic_radiance} yields the ballistic intensity $P_0(\mathbf{r_s},\mathbf{r},t)$:
\begin{equation}
P_0(\mathbf{r_s},\mathbf{r},t) =\frac{e^{-f(\mathbf{r_s},\mathbf{r})/l_{ext}}}{2 \pi |\mathbf{r}-\mathbf{r_s}|} \delta(|\mathbf{r}-\mathbf{r_s}|-ct)
\label{eqn:ballistic_intensity}
\end{equation}

Now let us consider the scattered contributions ($N \geq 1$). Similarly, the differential operators on the left hand side of Eq.\ref{eqn:partial_eq} can be integrated, to yield  
\begin{eqnarray}
& & P_N(\mathbf{r_s},\mathbf{r},t,\mathbf{u}) = \int_{0}^{\infty} dr_0 l_e^{-1} M(\mathbf{r_s}+r_0 \mathbf{u}) e^{-f(\mathbf{r_s},\mathbf{r_s}+r_0 \mathbf{u})/l_{ext}} P_{N-1}(\mathbf{r_s}+r_0 \mathbf{u},\mathbf{r},t-r_0/c) \nonumber \\
\label{eqn:MS_intensity_2}
& & P_N(\mathbf{r_s},\mathbf{r},t) = \frac{1}{l_{e}}\int \! \! \int d^2 \mathbf{r_1} M(\mathbf{r_1})\frac{e^{-f(\mathbf{r_s},\mathbf{r_1})/l_{ext}}}{2 \pi |\mathbf{r_1}-\mathbf{r_s}| } P_{N-1}(\mathbf{r_1},\mathbf{r} ,t-|\mathbf{r_1}-\mathbf{r_s}|/c)  
\end{eqnarray}
In this expression, the first scattering takes place in $\mathbf{r_1}$, somewhere inside $ (\vartheta )$. $e^{-f(\mathbf{r_s},\mathbf{r_1})/l_{ext}}/(2 \pi |\mathbf{r_1}-\mathbf{r_s}| )$ corresponds to the amplitude of the ballistic intensity at $\mathbf{r_1}$. The ratio $d^2 \mathbf{r_1}/l_e$ can be interpreted as a scattering cross section (actually a length, in 2D) of the infinitesimal surface element $d^2 \mathbf{r_1}$. The intensity propagation from $\mathbf{r_1}$ to $\mathbf{r}$ and the associated $N-1$ scattering events are taken into account by $P_{N-1}$. Hence the partial intensities can be obtained recursively.

From the expression of $P_0$ (Eq.\ref{eqn:ballistic_intensity}), we infer the single-scattering term $P_1$:
\begin{eqnarray}
P_1(\mathbf{r_s},\mathbf{r},t) & = &\frac{1}{l_{e}}\int \! \! \int d^2\mathbf{r_1} M(\mathbf{r_1}) \frac{e^{-f(\mathbf{r_s},\mathbf{r_1})/l_{ext}}}{2 \pi |\mathbf{r_1}-\mathbf{r_s}| } P_{0}(\mathbf{r_1}, \mathbf{r} ,t-|\mathbf{r_1}-\mathbf{r_s}|/c)  \nonumber \\
& =& \frac{1}{l_e} \int \! \! \int d^2 \mathbf{r_1} M(\mathbf{r_1}) \frac{e^{-f(\mathbf{r_s},\mathbf{r_1})/l_{ext}}}{2 \pi |\mathbf{r_1}-\mathbf{r_s}| } \frac{e^{-f(\mathbf{r_1},\mathbf{r})/l_{ext}}}{2 \pi |\mathbf{r}-\mathbf{r_1}|} \delta(|\mathbf{r_1}-\mathbf{r_s}| + |\mathbf{r}-\mathbf{r_1}|-ct)
\label{eqn:intensity_1}
\end{eqnarray}
This expression can be used to calculate the single scattering contribution $I^S$ for comparison with the experimental measurements. The transducers have a finite size $b$, hence a directivity pattern $\mathcal{D}(\theta)$:
\begin{equation}
\mathcal{D}(\theta)=\frac{\sin \left ( \pi b f_c \tan \theta /c \right)}{\pi b f_c \tan \theta /c }
\end{equation}
where $f_c$ is the central frequency. The same transducer is used as a source and a receiver. For convenience, it is placed at the origin: $\mathbf{r_s}=\mathbf{r}=\mathbf{0}$. Hence from Eq.\ref{eqn:intensity_1}, we have
\begin{equation}
\label{eqn:IS_1}
I^S(X=0,t)  = P_1(\mathbf{0},\mathbf{0},t) = \frac{1}{4 \pi^2 l_e} \int \! \! \int_{(\vartheta)} d \mathbf{r_1} \left | \mathcal{D}(\theta_1) \right |^4 \frac{e^{-2f(\mathbf{0},\mathbf{r_1})/l_{ext}}}{r_1^2 } \delta(2r_1-ct)
\end{equation}
Using polar coordinates $\mathbf{r_1}=(r_1,\theta_1)$ and with $f(\mathbf{0},\mathbf{r_1})=r_1 - \frac{a}{\cos \theta_1}$ (see Fig.\ref{fig:fig11}), we obtain:
\begin{equation}
\label{eqn:ss_4}
I^S(X=0,t)=\frac{e^{-ct/l_{ext}}}{4 \pi^2 l_e ct} \Theta(ct-2a) \int_{-\theta_{max}}^{\theta_{max}}d\theta_1 \left | \mathcal{D}(\theta_1) \right |^4 \exp \left (\frac{2a}{l_{ext}\ cos \theta_1 }\right) 
\end{equation} 
with $\theta_{max}= \arccos \left ( \frac{2 a}{ct} \right )$ {and $\Theta$ the Heaviside function}. 
Eq.\ref{eqn:ss_4} is the simplest analytical form that we can obtain for the single-scattering intensity in our configuration. $I^S(t)$ can only be computed numerically (see Fig.6(a) of the manuscript). Once it is normalized by its maximum, the evolution of the single scattering intensity with time only depends on the extinction length $l_{ext}$.
\begin{figure}[htbp] 
\includegraphics{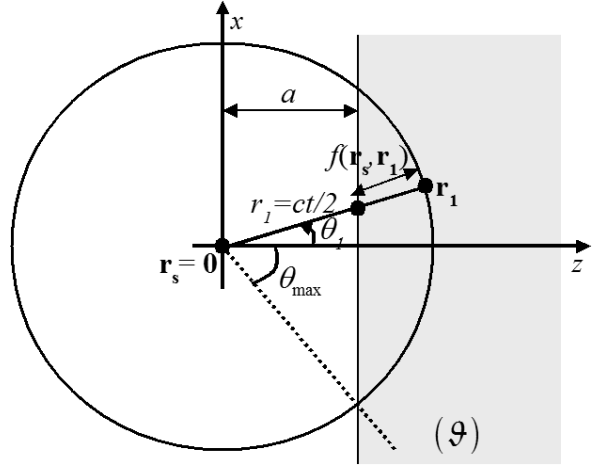}
\caption{\label{fig:fig11}Typical configuration used for the analytical calculation of the single scattering intensity $I^S(X=0,t)$}
\end{figure}

We now turn to higher orders of scattering, starting with $N=2$. Combining Eq.\ref{eqn:MS_intensity_2} and \ref{eqn:intensity_1} , we obtain:
\begin{eqnarray}
P_2(\mathbf{r_s},\mathbf{r},t)  &=& \frac{1}{l_{e}^2} \int \! \! \int d^2 \mathbf{r_1} M(\mathbf{r_1}) \frac{e^{-f(\mathbf{r_s},\mathbf{r_1})/l_{ext}}}{2 \pi |\mathbf{r_1}-\mathbf{r_s}| } \int \! \! \int d^2\mathbf{r_2} M(\mathbf{r_2}) \nonumber \\
& \times & \frac{e^{-f(\mathbf{r_2},\mathbf{r})/l_{ext}} e^{-f(\mathbf{r_1},\mathbf{r_2})/l_{ext}}}{2 \pi |\mathbf{r_2}-\mathbf{r_1}|2 \pi |\mathbf{r}-\mathbf{r_2}| } \delta(|\mathbf{r_1}-\mathbf{r_s}| + |\mathbf{r_2}-\mathbf{r_1}|+|\mathbf{r}-\mathbf{r_2}|-ct)
\label{eqn:intensity_2b}  
\end{eqnarray}
Rigorously, we would have to calculate all partial intensities of order N to obtain
a theoretical expression of the multiple scattering intensity $P^M(\mathbf{r_s},\mathbf{r},t)=\sum_{N=2}^\infty P_i(\mathbf{r_s},\mathbf{r},t)$.
This would be particularly tedious, time-consuming numerically, and useless in our experimental case. A simplified way is proposed to approach the multiple scattering intensity. Summation of Eqs.\ref{eqn:ballistic_intensity}-\ref{eqn:MS_intensity_2} over all $N$ results in:
\begin{eqnarray}
& & P(\mathbf{r_s},\mathbf{r},t) =  P_0(\mathbf{r_s},\mathbf{r},t)+ \int \! \! \int d^2 \mathbf{r_1}  \frac{M(\mathbf{r_1}) e^{-f(\mathbf{r_s},\mathbf{r_1})/l_{ext}}}{2 \pi l_e |\mathbf{r_1}-\mathbf{r_s}| } P(\mathbf{r_1},\mathbf{r} ,t-|\mathbf{r_1}-\mathbf{r_s}|/c) \label{eqn:mord_la_queue} \\
& & = P_0(\mathbf{r_s},\mathbf{r},t) + \underbrace{\int \! \! \int d^2 \mathbf{r_1}  \frac{M(\mathbf{r_1})e^{-f(\mathbf{r_s},\mathbf{r_1})/l_{ext}}}{2 \pi l_e |\mathbf{r_1}-\mathbf{r_s}| } P_0(\mathbf{r_1},\mathbf{r} ,t-|\mathbf{r_1}-\mathbf{r_s}|/c)}_{P_1(\mathbf{r_s},\mathbf{r},t) }\nonumber \\
  & &+  \underbrace { \int \! \! \int d^2 \mathbf{r_1}  \frac{M(\mathbf{r_1}) e^{-f(\mathbf{r_s},\mathbf{r_1})/l_{ext}}}{2 \pi l_e  |\mathbf{r_1}-\mathbf{r_s}| } \left [ P(\mathbf{r_1},\mathbf{r} ,t-|\mathbf{r_1}-\mathbf{r_s}|/c)-P_0(\mathbf{r_1},\mathbf{r} ,t-|\mathbf{r_1}-\mathbf{r_s}|/c) \right]}_{P^M(\mathbf{r_s},\mathbf{r},t)}
  \label{eqn:P^M}
\end{eqnarray}
The expression of $P^M$ (Eq.\ref{eqn:P^M}) can be rewritten using the spatial reciprocity theorem. The vector positions $\mathbf{r}$ and $\mathbf{r_1}$ can be exchanged in $P(\mathbf{r_1},\mathbf{r} ,t-|\mathbf{r_1}-\mathbf{r_s}|/c)$.
$P^M$ then becomes
\begin{eqnarray}
P^M(\mathbf{r_s},\mathbf{r},t) = \frac{1}{l_{e}}\int \! \! \int d^2 \mathbf{r_1} & & M(\mathbf{r_1})  \frac{e^{-f(\mathbf{r_s},\mathbf{r_1})/l_{ext}}}{2 \pi |\mathbf{r_1}-\mathbf{r_s}| } \nonumber \\
&  & \times \left [ P(\mathbf{r},\mathbf{r_1} ,t-|\mathbf{r_1}-\mathbf{r_s}|/c)-P_0(\mathbf{r},\mathbf{r_1} ,t-|\mathbf{r_1}-\mathbf{r_s}|/c) \right]
  \label{eqn:P^M2}
\end{eqnarray}
Replacing $P(\mathbf{r},\mathbf{r_1} ,t-|\mathbf{r_1}-\mathbf{r_s}|/c)$ by its expression in Eq.\ref{eqn:mord_la_queue} finally leads to:
\begin{eqnarray}
P^M(\mathbf{r_s},\mathbf{r},t) &=& \frac{1}{l_{e}^2}\int \! \! \int d^2 \mathbf{r_1} M(\mathbf{r_1}) \frac{e^{-f(\mathbf{r_s},\mathbf{r_1})/l_{ext}}}{2 \pi |\mathbf{r_1}-\mathbf{r_s}| } \nonumber \\
& \times & \int \! \! \int d^2 \mathbf{r_2} M(\mathbf{r_2}) \frac{e^{-f(\mathbf{r},\mathbf{r_2})/l_{ext}}}{2 \pi |\mathbf{r}-\mathbf{r_2}| }  
 P(\mathbf{r_1},\mathbf{r_2} ,t-(|\mathbf{r_1}-\mathbf{r_s}|+|\mathbf{r}-\mathbf{r_2}|)/c)
  \label{eqn:P^M3}
\end{eqnarray}
In this expression of $P^M$, $\mathbf{r_1}$ and $\mathbf{r_2}$ can be interpreted as the points where the first and the last scattering events take place. The propagation of the intensity between $\mathbf{r_1}$ and $\mathbf{r_2}$ is described by $P$. Eq.\ref{eqn:P^M3} is still implicit since $P^M$ is actually contained in $P$. An approximate solution for Eq\ref{eqn:P^M3} can be obtained by taking for $P$ the solution of the RTE for the case of an infinite random medium \cite{paasschens}
\begin{eqnarray}
P_{\infty}(\mathbf{r_1},\mathbf{r_2} ,t) & = & P_{\infty}(r=|\mathbf{r_2}-\mathbf{r_1}|,t)\nonumber \\
& = &\underbrace{\frac{e^{-ct/l_{ext}}}{2\pi r}\delta(ct-r)}_{\mbox{ballistic term}} \nonumber \\
& + & \underbrace{\frac{e^{-ct/l_{ext}}}{2\pi l_e c t}\left ( 1-\frac{r^2}{c^2 t^2}\right )^{-1/2} \exp \left ( \frac{\sqrt{c^2t^2-r^2}}{l_e}\right ) \Theta (ct-r)}_{\mbox{scattering term}}
\label{eqn:TRE}
\end{eqnarray}
Replacing $P$ by $P_{\infty}$ in Eq.\ref{eqn:P^M3} allows to approach the multiple scattering intensity $P^M$:
 \begin{eqnarray}
P^M(\mathbf{r_s},\mathbf{r},t)  &\simeq & \frac{1}{l_{e}^2}\int \! \! \int d^2 \mathbf{r_1} M(\mathbf{r_1}) \frac{e^{-f(\mathbf{r_s},\mathbf{r_1})/l_{ext}}}{2 \pi |\mathbf{r_1}-\mathbf{r_s}| } \nonumber \\
& \times & \int \! \! \int d^2 \mathbf{r_2} M(\mathbf{r_2}) \frac{e^{-f(\mathbf{r},\mathbf{r_2})/l_{ext}}}{2 \pi |\mathbf{r}-\mathbf{r_2}| } 
 P_{\infty}(\mathbf{r_1},\mathbf{r_2} ,t-(|\mathbf{r_1}-\mathbf{r_s}|+|\mathbf{r}-\mathbf{r_2}|)/c)
  \label{eqn:P_Maj}
\end{eqnarray}
One can object that it would have been simpler to replace $P$ by $P_{\infty}$ in Eq.\ref{eqn:P^M}. Yet this would amount to considering scattering paths whose last scatterer is outside of $(\vartheta)$ which is unphysical. As a consequence, the double scattering contribution would have been largely overestimated. On the contrary, Eq.\ref{eqn:P_Maj} only takes into account scattering paths for which the first and last scatterers are within the random medium $(\vartheta)$. Therefore Eq.\ref{eqn:P_Maj} exactly accounts for the double scattering contribution. Yet it overestimates higher orders of scattering. 

The directivity patterns of the transmitting/receiving elements as well as the distance $a$ between the array and the sample can be taken into account in the calculation of the multiple scattering intensity, in a similar manner as what was done for the single scattering contribution. At exact backscattering ($X=0$), because of the coherent backscattering effect \cite{akkermans3}, the multiple scattered intensity is twice that predicted from RTE. Hence we have, for the double scattering intensity :
\begin{eqnarray}
 I^{(2)}(X=0,t) & =&  2 P_2(0,0,t) \nonumber \\  &=& \frac{2}{l_{e}^2} \int \! \! \int_{(\vartheta)} d^2 \mathbf{r_1}  \frac{\left | \mathcal{D}(\theta_1) \right |^2 e^{-f(\mathbf{0},\mathbf{r_1})/l_{ext}}}{2 \pi r_1}  \nonumber \\
& \times & \int \! \! \int_{(\vartheta)} d^2\mathbf{r_2} 
 \frac{\left | \mathcal{D}(\theta_2) \right |^2 e^{- ( f(\mathbf{r_2},\mathbf{0})+f(\mathbf{r_1},\mathbf{r_2}))/l_{ext}}}{ 4 \pi^2 |\mathbf{r_2}-\mathbf{r_1}| r_2 } \delta(r_1 + |\mathbf{r_2}-\mathbf{r_1}|+r_2-ct)
\label{eqn:intensity_inc_2}  
\end{eqnarray}
with $f(\mathbf{0},\mathbf{r_i})=r_i-a/\cos \theta_i$ and $f(\mathbf{r_1},\mathbf{r_2})= |\mathbf{r_2}-\mathbf{r_1}|$. 
\begin{figure}[htbp] 
\includegraphics{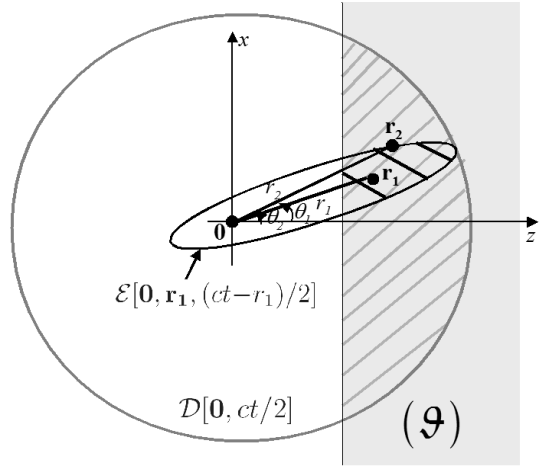}
\caption{\label{fig:fig12}Typical configuration used for the analytical calculation of the double scattering intensity $I^{(2)}(0,t)$}
\end{figure}

The Dirac function which appears upon the integrand of Eq.\ref{eqn:intensity_inc_2} implies that the integral over $\mathbf{r_1}$ has to be performed over the intersection between the disk $\mathcal{D}[\mathbf{0}, ct/2]$ (whose center is $\mathbf{0}$ and radius is $ct/2$) and the random medium $(\vartheta)$ (the grey cross hatched surface in Fig.\ref{fig:fig12}). The integration over $\mathbf{r_2}$ is performed along the intersection between the ellipse $\mathcal{E}[\mathbf{0},\mathbf{r_1}, (ct-r_1)/2 ]$ (whose foci are $\mathbf{0}$ and $\mathbf{r_1}$, and semi major axis is $(ct-r_1)/2 $) and the random medium $(\vartheta)$ (see Fig.\ref{fig:fig12}). Using polar coordinates, Eq.\ref{eqn:intensity_inc_2} becomes
\begin{eqnarray}
I^{(2)}(X=0,t) & = & \frac{e^{-ct/l_{ext}}}{ 4 \pi^3 l_{e}^2} \int \! \! \int_{\mathcal{D}\bigcap (\vartheta)} d r_1 d \theta_1  \left | \mathcal{D}(\theta_1) \right |^2 e^{a/(l_{ext}\cos \theta_1)}  \nonumber \\
& \times & \int_{\theta_{2,min}}^{\theta_{2,max}} \!\!\!\!\!\! d \theta_2  \left | \mathcal{D}(\theta_2) \right |^2  \frac{2 e^{a/(l_{ext}\cos \theta_2)} \left[ (ct -r_1(1+ \cos(\theta_2 -\theta_1))  \right]}{(ct -2r_1)^2+4(ct -r_1)r_1 \sin^2 \left (\frac{\theta_2 - \theta_1}{2} \right) + r_1^2}
\label{eqn:intensity_inc_2b}  
\end{eqnarray}
Eq.\ref{eqn:intensity_inc_2b} is the simplest analytical form for the double-scattering intensity in our configuration. $I^{(2)}$ can be computed numerically (see Fig.6(b) of the manuscript).

Finally, we consider the approximate solution for the multiple scattering intensity (Eq.\ref{eqn:P_Maj}), and take into account the coherent backscattering phenomenon, as well as the directivity of the sources/receivers to obtain
 \begin{eqnarray}
I^{M}(X=0,t) & \simeq & \frac{2}{l_{e}^2}\int \! \! \int_{(\vartheta)} d^2 \mathbf{r_1} \left | \mathcal{D}(\theta_1) \right |^2 \frac{e^{-f(\mathbf{0},\mathbf{r_1})/l_{ext}}}{2 \pi r_1 } \nonumber \\ 
& \times & \int \! \! \int_{(\vartheta)} d^2 \mathbf{r_2}\left | \mathcal{D}(\theta_2) \right |^2 \frac{e^{-f(\mathbf{0},\mathbf{r_2})/l_{ext}}}{2 \pi r_2 } P_{\infty}(|\mathbf{r_2}-\mathbf{r_1}| ,t-(r_1+r_2)/c)
  \label{eqn:incoherent_intensity}
\end{eqnarray}
Using polar coordinates, Eq.\ref{eqn:incoherent_intensity} becomes
 \begin{eqnarray}
I^{M}(X=0,t)  &\simeq & \frac{1}{2 \pi ^2l_{e}^2}\int \! \! \int_{\mathcal{D}\bigcap (\vartheta)} d r_1 d\theta_1 \left | \mathcal{D}(\theta_1) \right |^2 e^{-(r_1-a/\cos \theta_1)/l_{ext}} \nonumber \\
&\times& \int \! \! \int_{\mathcal{E}\bigcap{(\vartheta)}} \! \!\! \!\! \!\! \!\! \!\! \! d r_2 d\theta_2 \left | \mathcal{D}(\theta_2) \right |^2 e^{-(r_2-a/\cos \theta_2)/l_{ext}} P_{\infty}(|\mathbf{r_2}-\mathbf{r_1}| ,t-(r_1+r_2)/c)
  \label{eqn:incoherent_intensity2}
\end{eqnarray}
Unlike Eqs.\ref{eqn:ss_4} and \ref{eqn:intensity_inc_2b}, Eq.\ref{eqn:incoherent_intensity2} is an approximate solution of the radiative transfer equation. It is the simplest analytical expression that we can obtain. 
It allows to compute the time-evolution of $I^M$ numerically (see Fig.6(b)). Note that $I^M(t)$  does not only depend on $l_{ext}$, but also on the relative weight of the diffusion ($l_e$) and the absorption ($l_a$) losses. This dependence on $l_e$ and $l_a$ is contained implicitly in the intensity propagator $P_{\infty}$(see Eq.\ref{eqn:TRE}).

\section{\label{app:app5}}

This appendix describes the numerical process we use to determine the upper bound $\lambda_{\mbox{\small{max}}}^{(0)}$ of the singular values distribution, when the variance of the element $a_{ij}$ is not the same for every line of the matrix $\mathbf{A}$.

The first step is to estimate this variance. This is done by measuring the mean intensity $I_i$ of $a_{ij}$ along each line $i$ of $\mathbf{A}$:
\begin{equation}
I_i  =  \left < \left | a_{i,j}\right |^2 \right >_{j}
\end{equation}
where the symbol $<. >$ denotes an average over $j$.

The second step consists in building a diagonal matrix $\Sigma$ of size $M \times M$, whose elements are:
\begin{equation}
\sigma_{ii}=  \sqrt{I_i}
\end{equation}

The next step is the generation of a random matrix $\mathbf{Q}$ with the same variance as $\mathbf{A}$. To that aim, a random matrix $\mathbf{P}$ of dimension $M \times L$ is first generated numerically. Its coefficients are independent and identically distributed. Then $\mathbf{Q}$ is built from $\mathbf{P}$ according to :
\begin{equation}
\mathbf{Q}=\mathbf{\Sigma}\mathbf{C}^{\frac{1}{2}}\mathbf{P}\mathbf{D}^{\frac{1}{2}} 
\end{equation}

This ensures that the $q_{ij}$ have the same variance and correlation properties than the $a_{ij}$. Once the SVD of $\mathbf{Q}$ is performed, its first singular value $\lambda_1$ is renormalized according to Eq.10. The same operation is repeated for 500 realisations, which yields a histogram of $\tilde{\lambda}_1$ and thereby an estimate of the distribution function $F_1$ of the first singular value, $F_1(\lambda)=\mbox{Prob} \{ \tilde{\lambda}_1 < \lambda\}$, can be estimated. The upper bound $\lambda_{\mbox{\small{max}}}^{(0)}$ is then deduced from $F_1$ (see Appendix \ref{app:app3}).

\end{document}